\documentclass[twocolumn,astrosymb]{aastex631}

\usepackage{textcomp}
\usepackage{amssymb}
\usepackage{amsmath}
\usepackage{comment}
\usepackage{wasysym}
\usepackage{CJK}
\usepackage{verbatim}
\usepackage{enumerate}

\shorttitle{Outflow in NGC 5195}
\shortauthors{Xu \& Wang}

\begin{document}

\begin{CJK*}{UTF8}{gbsn}
\title{Ghost in the Shell: Evidence for Past AGN Activities in NGC 5195 \\ from a Newly Discovered large-scale Ionized Structure}

\correspondingauthor{Junfeng Wang}
\email{jfwang@xmu.edu.cn}

\author[0000-0003-0970-535X]{Xiaoyu Xu (许啸宇)}
\affiliation{Department of Astronomy, Xiamen University, Xiamen, Fujian 361005, China}

\author[0000-0003-4874-0369]{Junfeng Wang}
\affiliation{Department of Astronomy, Xiamen University, Xiamen, Fujian 361005, China}

\begin{abstract}

The early-type galaxy NGC 5195 (alternatively known as M51b) possesses extended gas features detected in multi-wavelength, postulated to be associated with previous activities of the central supermassive black hole (SMBH). Using integral field spectroscopic observations from the Canada-France-Hawaii Telescope (CFHT)/SITELLE, we report on the discovery of a new large-scale ionized gas structure traced by [O~{\sc{iii}}], [N~{\sc{ii}}], and H$\alpha$ line emission, extending to $\sim10\rm\,kpc$ from the nucleus of NGC 5195. 
Its bipolar morphology, emission line ratio diagnostics, and comparison with the X-ray image from {\em Chandra} and low-frequency radio data from {\em LOFAR} all indicate that it is likely an outflow inflated by a past episode of elevated active galactic nucleus (AGN) activity.
Assuming the ionized gas is outflowing from the central region of NGC 5195, the estimated mass and energy outflow rates are $\dot{M}_{\rm out} = 3.5$--$27.9 \rm\, M_{\astrosun}\, yr^{-1}$ and $\dot{E}_{\rm out} = 0.98$--$7.9\times10^{40}\rm\, erg\, s^{-1}$, respectively, which cannot be provided by current star formation and the low luminosity nucleus.  
Alternatively, considering the history of gravitational interaction between the M51 pair and the presence of HI tidal tail, the northern large-scale ionized gas could very likely be associated with tidally stripped material illuminated by a luminous AGN in the past.

\end{abstract}

\keywords{Interacting galaxies(802) --- LINER galaxies(925) --- Interstellar medium(847) --- Circumgalactic medium(1879)}

\section{Introduction} \label{sec:intro}

Galaxy interactions and mergers are thought to play an important role in triggering star formation and active galactic nuclei (AGNs) activities.  
Galactic outflows driven by AGNs are predicted by theoretical models and simulations during such a key stage of galaxy evolution. 
Unequivocally, multiphase AGN outflows on various scales have been detected in galaxies at different redshifts \citep[e.g.][]{2012ARA&A..50..455F,2015Natur.519..436T,2016ApJ...817..108W,2018A&A...619A..74V}. 
Such AGN activities are postulated to be episodic, possibly governed by accretion disk physics and large-scale gas supply \citep{2009ApJ...690...20S,2021ApJ...921...70S}. 
The impact on the surrounding galactic medium leaves imprints that provide useful diagnostics on the historic radiative and mechanical power of these AGNs \citep[e.g.,][]{2013ApJ...773..148S,2016MNRAS.457.3629S,2017ApJ...835..256K,2022arXiv220805915H}.

The interacting system Messier 51 (M51) contains a grand design spiral galaxy NGC 5194 (M51a) and an early-type galaxy NGC 5195 (M51b).
NGC 5195 is classified as different types of the galaxy, such as SB$0_{1}$-pec \citep[][]{1981rsac.book.....S} and I0-pec \citep[][]{1991rc3..book.....D}, due to its peculiar morphology that is tidally disturbed in the interaction.
An SMBH with mass $M_{\rm BH}\sim 10^{7}\rm\,M_{\astrosun}$ is believed to reside in the nucleus of NGC 5195 \citep[e.g.][]{2016ApJ...823...75S,2018MNRAS.476.2876R}.
According to the optical emission-line ratios, the nucleus is classified as the low ionization nuclear emission-line region (LINER) \citep[][]{1997ApJS..112..315H}.
An AGN is suggested to inhabit the center of this galaxy based on the detection of the high-excitation emission line [Ne~{\sc{v}}]$\lambda 14.32\rm\,\mu m$ \citep[][]{2009MNRAS.398.1165G}.
The current level of the nuclear activity of NGC 5195 is expected to be very low with an unabsorbed X-ray luminosity of $L_{\rm 0.3-10\,keV}\sim 1.3\times10^{38}\rm\, erg\,s^{-1}$ and weak radio emission of $L_{\rm 8.6\,GHz}<8.7\times10^{34}\rm\, erg\,s^{-1}$ \citep[e.g.][]{2018MNRAS.476.2876R}.

\cite{2016ApJ...823...75S} revealed a double-arc-like X-ray structure in the southern region of the NGC 5195 nucleus with $\sim 15\arcsec$--$30\arcsec$ away from the center using the \textit{Chandra} data.
A slender H$\alpha$-emitting feature \citep[first detected by][]{1998ApJ...506..135G} locates just outside the outer X-ray arc, which was expected to be plowed up by the X-ray-emitting gaseous outflow.
They attributed the arcs to episodic outbursts of the nuclear SMBH.
Later \cite{2018MNRAS.476.2876R} confirmed the kpc-scale arcs in the radio bands with Very Large Array (VLA) observations and detected a linear radio spur near the nucleus pointing towards the arcs.  
They suggested that the kpc-scale structures likely resemble a jet-inflated bubble associated with past AGN activities.  Additionally, a vast H$\alpha$ cloud $\sim13\arcmin$ north of the M51 system was detected by \cite{2018ApJ...858L..16W}, which was expected to be associated with the tidal stripping or AGN activities during the interaction.

In this work, we present the results of our new integral field spectroscopic (IFS) observations focusing on the large ionized gas structure in NGC 5195.
The observations and data reduction are described in Section~\ref{sec:data}.  Main results including the flux and velocity maps, the emission-line diagnostic diagram and map, and the comparison with other observational results in different bands are presented in Section~\ref{sec:result}.
In Section~\ref{sec:discussion} and Section~\ref{sec:summary}, we discuss the findings and draw brief conclusions.
For consistency with \cite{2016ApJ...823...75S} and \cite{2018MNRAS.476.2876R}, the distance to NGC 5195 is adopted as $D=8.0 \rm\,Mpc$ ($1\arcsec \approx 40\rm\, pc$).

\section{Observations and Data Reduction} \label{sec:data}

\subsection{CFHT/SITELLE data} \label{sec:sitelle}

NGC 5195 was observed by the instrument SITELLE \citep{2012SPIE.8446E..0UG} installed on the 3.6m CFHT, on 2019-04-08, 2019-04-10, and 2019-05-10 (Program 19AS002, PI: J.~Wang). 
SITELLE is an imaging Fourier transform spectrometer (IFTS) designed for the CFHT, which can provide IFS abilities in the visible band (350 to 900 nm), with a large field of view (FOV) $\sim 11\times 11$ arcmin$^{2}$ and variable spectral resolution (R = 2 to 10000) \citep{2019MNRAS.485.3930D}. 
We used the SN2 (480-520 nm) and SN3 (651-685 nm) filters, with exposure time $\sim$4.5 hours for SN2 and $\sim$3.55 hours for SN3, to cover main emission lines e.g. H$\alpha$, H$\beta$, [O~{\sc{iii}}]$\lambda\lambda 4959, 5007$, [N~{\sc{ii}}]$\lambda\lambda$6548,6583, and [S~{\sc{ii}}]$\lambda\lambda$6716,6731. 
The spectral resolution of the SN2 and SN3 datacubes is set to be $R\sim 1150$. 

The calibrated datacubes of SN2 and SN3 were retrieved from the Canadian Astronomy Data Centre (CADC)\footnote{https://www.cadc-ccda.hia-iha.nrc-cnrc.gc.ca/en/}. 
The ORCS\footnote{https://github.com/thomasorb/orcs} \citep[][]{2015ASPC..495..327M} is used to analyse the data, which is a python module for analysis of SITELLE spectral cubes. 
To obtain extended emission lines with enough signal-to-noise ratio (SNR), we rebinned the spectral cubes with a factor of 10.
The Voronoi binning technique \citep[][]{2003MNRAS.342..345C} was also used to rebin the datacubes for cross-checking the faint extended structure.  
Comparing the results of these two binning methods, no significant improvements can be gained using the Voronoi binning, hence the 10$\times$10 binning is sufficient for the main purpose of this work.
We follow the method described by \cite{2018MNRAS.477.4152R} to subtract the stellar continuum from the spectra.
The spectrum without emission lines extracted from the nuclear region with a radius of $\sim 5.1\arcsec \times 5.1\arcsec$ is used as the continuum template, which is subtracted in every spectrum extracted from the FOV after normalized to the continuum levels of those spectra.
Due to the low spectral resolution, we only used the ``sinc" model, the natural instrument line shape of IFTS, to fit emission lines of the SN2 and SN3 datacube and thus we cannot obtain the velocity dispersion of emission lines.
We followed the procedure described in \cite{2018MNRAS.473.4130M} to refine the wavelength calibration by measuring the velocity of the night sky emission lines.

For the extinction correction, the flux ratio of H$\alpha$ and H$\beta$ (the Balmer decrement) was used. The attenuation law from \cite{2001PASP..113.1449C} and an intrinsic Balmer decrement $\rm H\alpha / H\beta=3.1$ \citep[][]{2006agna.book.....O} were adopted.

\begin{figure*}[ht!]
	\includegraphics[width=\textwidth]{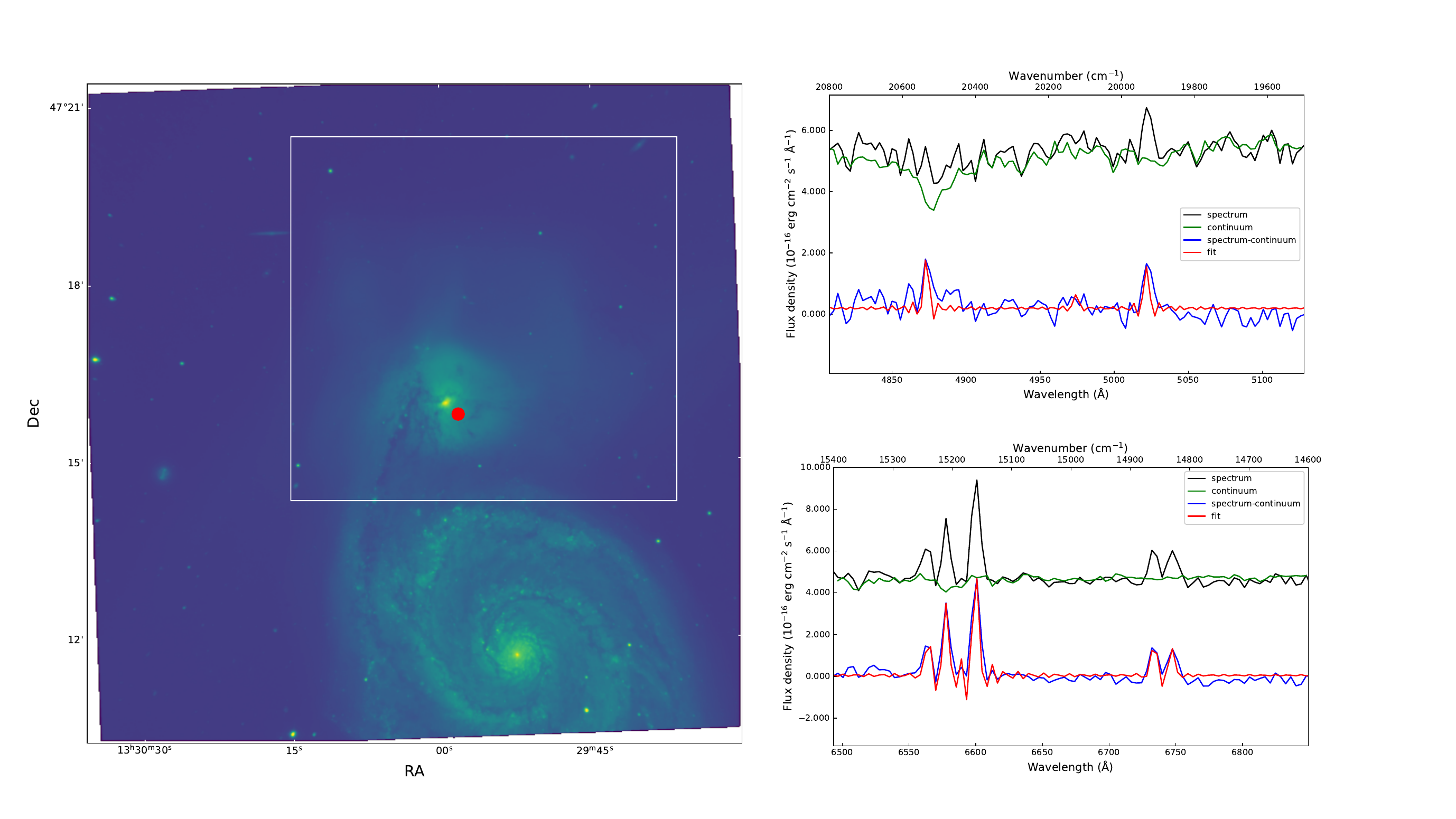}
    \caption{Left: The deep broad-band image of M51 obtained from SITELLE's SN2 datacube. 
    The large spiral galaxy in the south is the NGC 5194 and the NGC 5195 is the smaller one above the NGC 5194. 
    The white box denotes the region for fitting and analysis. North is to the top, and east to the left.
    Top right: A spectrum (black) with the continuum template (green), emission-line residual (blue), and fitting model (red) extracted from the red circle region in the left panel of the SN2 datacube. 
    Bottom right: Same as the top right but extracted from the SN3 datacube. }
    \label{fig:df2_spec}
\end{figure*}

Figure~\ref{fig:df2_spec} shows the deep image derived from the SN2 datacube. The region covering NGC 5195 is selected for fitting and analysis, shown as a white box.  
A large circle region far away from the M51 system and including no apparent point sources is selected to extract a spectrum as the background. 
Two examples of spectra extracted from the red circular region in the left panel of Figure~\ref{fig:df2_spec} are shown in the right panels, in which the top is for the SN2 and the bottom is for the SN3 datacube. 
The spectral fitting illustrates that the continuum templates perform well especially for the $\rm H\beta$ emission line fitting.

\subsection{{\em Chandra} X-ray data} \label{sec:xray}

The X-ray emission of the NGC 5195 has been thoroughly studied in detail by \cite{2016ApJ...823...75S} and \cite{2018MNRAS.476.2876R}.
This work would not repeat their analysis but only derive the soft X-ray image to compare with our IFS results.
We obtained all available {\em Chandra} observations of M51 from {\em Chandra} Data Archive\footnote{\url{https://cda.harvard.edu/chaser/}} except ObsID 414 and ObsID 12562 \citep[for the reason described in][]{2016ApJ...823...75S}, including all data used by \cite{2016ApJ...823...75S} and \cite{2018MNRAS.476.2876R} supplemented with ObsID 20998, 23472, 23473, 23474, and 23475.
The total exposure time reaches 1.04 Ms. All data were reprocessed and merged using CIAO 4.13 \citep[][]{2006SPIE.6270E..1VF} with CALDB version 4.9.4.

\section{Results} \label{sec:result}

\subsection{Flux and velocity maps} \label{subsec:flux-vel-maps}

\begin{figure*}
\includegraphics[width=\textwidth,trim=0 90 0 90]{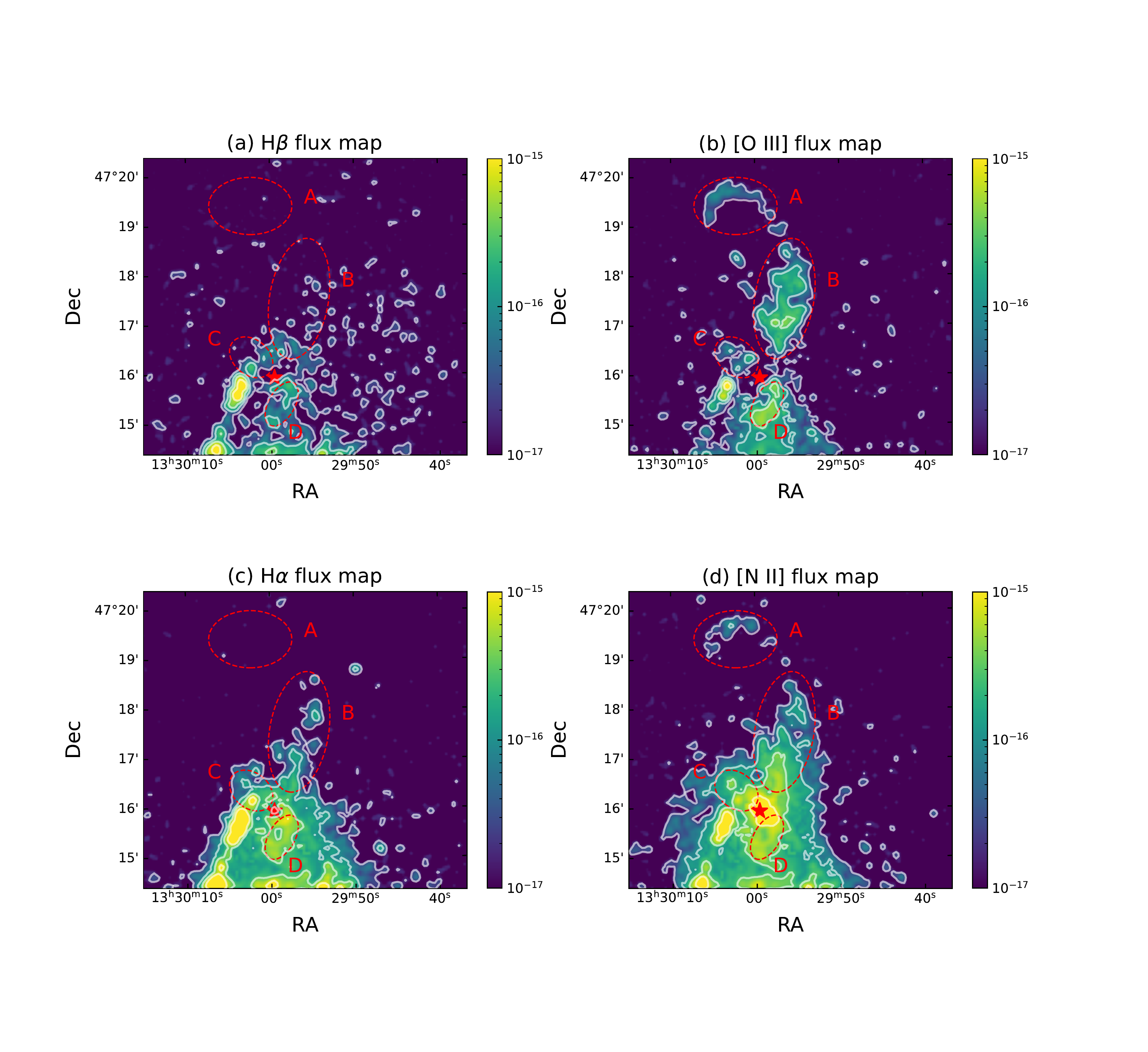}
\caption{(a): $\rm 10 \times 10$ binned and smoothed (Gaussian kernel with radius of 1 spaxel) flux map of H$\beta$ emission line superposed with its flux contours. 
The contour levels are $2 \times 10^{-17}$, $1\times 10^{-16}$, $3\times10^{-16}$, and $9\times10^{-16}$. 
The unit of the colorbar is $\rm erg\, cm^{-2}\, s^{-1}$. 
The spaxels with SNR$<3$ are rejected. 
(b), (c), and (d) are the same as (a) but for [O~{\sc{iii}}]$\lambda5007$, H$\alpha$, and [N~{\sc{ii}}]$\lambda6583$, respectively.
The red star marks the NGC 5195 nucleus. Four red dashed elliptical regions are labeled as A, B, C, and D, respectively.
}
\label{fig:fmap}
\end{figure*}

In Figure~\ref{fig:fmap}, the rebinned flux maps of H$\beta$, [O~{\sc{iii}}]$\lambda5007$, H$\alpha$, and [N~{\sc{ii}}]$\lambda6583$ are shown. 
Due to the low SNRs of [S~{\sc{ii}}]$\lambda\lambda$6716,6731 in most spaxels, we do not show the flux maps of these two emission lines.
To obtain robust measurements, the spaxels with SNR$<3$ are rejected. 
Note that the lowest level of flux contours shown corresponds to $2 \times 10^{-17} \rm\, erg\, cm^{-2}\, s^{-1}$, and emission structures fainter than this threshold are not reliable, which can be affected by the instrumental modulation of SITELLE.
The most intriguing structure is the large hook-like [O~{\sc{iii}}] cloud (region A and B in Figure~\ref{fig:fmap}(b)) in the north of NGC 5195, extending $\sim4\arcmin$ in projection ($\sim10 \rm\, kpc$) from the nucleus, which is not known from previous works. 
Its presence is further confirmed in the [N~{\sc{ii}}] flux map.  
The H$\alpha$ flux maps also show extended features in the region B along the direction of the [O~{\sc{iii}}] hook with $\sim3\arcmin$ away from the nucleus.
This ionized gas structure is consistent with a similar faint bubble-like H$\alpha$ feature detected by \cite{2018ApJ...858L..16W} using deep narrow-band imaging.
Although the H$\beta$ flux map exhibit relatively weak emission, the main features are basically consistent with other flux maps.

In previous studies, a double arc-like structure in the south region of NGC 5195 had been detected in the soft X-ray band \citep[][]{2016ApJ...823...75S}, and corresponding structures are detected in H$\alpha$ narrowband image and $20\rm\,cm$ radio band \citep{1998ApJ...506..135G,2016ApJ...823...75S,2018MNRAS.476.2876R}. 
These structures can also be found in our emission-line flux maps, e.g. a blurred arc feature in H$\alpha$, [O~{\sc{iii}}], and [N~{\sc{ii}}] map (region D), which is extending to $\sim70\arcsec$ (for [O~{\sc{iii}}]) in projection. 

[O~{\sc{iii}}] emission appears to be very weak or absent in the nuclear region.
The dust extinction in the nuclear region of NGC 5195 is up to $A_{\rm V}=1.67\rm\, mag$ \citep[][]{2021RAA....21....6W}, which can reduce the observed flux of the [O~{\sc{iii}}] emission line.
Considering the low X-ray (an absorption corrected luminosity $L_{\rm 0.3-10\,keV}\sim 1.3\times10^{38}\rm\, erg\,s^{-1}$) and radio luminosity ($L_{\rm 8.6\,GHz}<8.7\times10^{34}\rm\, erg\,s^{-1}$) in the nuclear region of NGC 5195 \citep[e.g.][]{2018MNRAS.476.2876R}, the nuclear [O~{\sc{iii}}] emission is considered to be intrinsically very weak.

In all these flux maps, a spiral arm from the NGC 5194 connecting with region C can be seen in the southeastern region of the NGC 5195's nucleus.
A low flux gap can be seen in the bottom of the FoV and under region D in all flux maps, which is suggestive that the gap separates the ionized gas emission from NGC 5195 (above the gap) and NGC 5194 (under the gap).

\begin{figure*}[ht!]
\includegraphics[width=1.1\textwidth,trim=70 0 70 0]{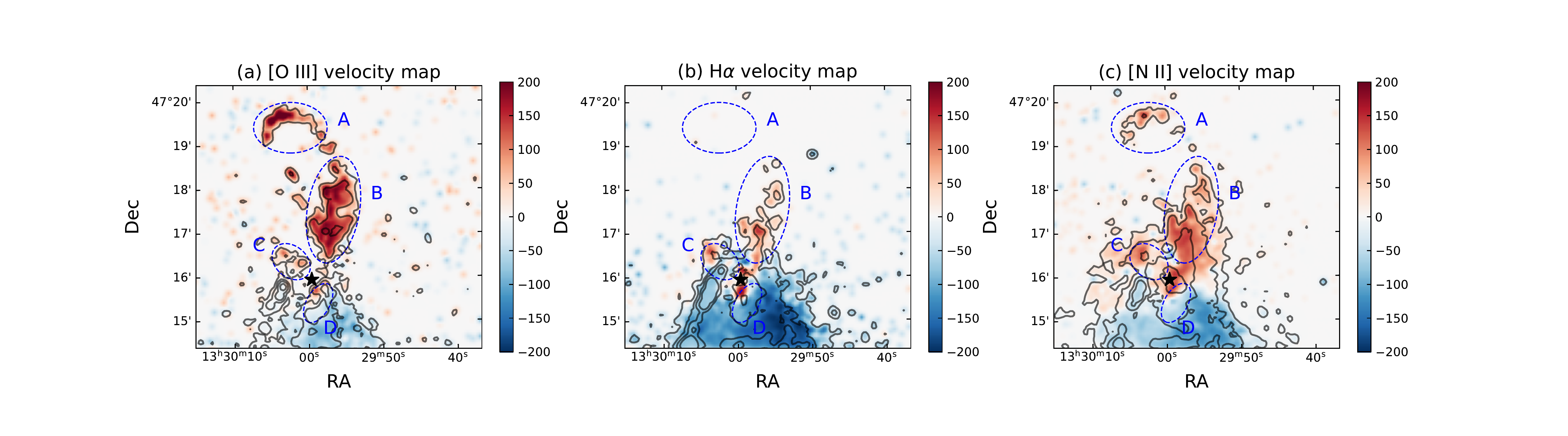}
\caption{(a): $\rm 10 \times 10$ binned and smoothed (Gaussian kernel with a radius of 1 spaxel) velocity map of [O~{\sc{iii}}]$\lambda5007$ superposed with the flux contours (same as contours in Figure~\ref{fig:fmap}).
(b) and (c) are the same as (a) but for H$\alpha$ and [N~{\sc{ii}}]$\lambda6583$, respectively.
The black star marks the position of the NGC 5195 nucleus and the four dashed regions are the same as those in Figure~\ref{fig:fmap}.
}
\label{fig:vmap}
\end{figure*}

In Figure~\ref{fig:vmap}, the velocity maps of [O~{\sc{iii}}], H$\alpha$, and [N~{\sc{ii}}] are presented.  
The systemic velocity $571\,\rm km\, s^{-1}$ of NGC 5195 \citep{1999PASP..111..438F} is used to calibrate these velocity maps.
As the [O~{\sc{iii}}] velocity map shows, the velocities in regions A, B, and C are redshifted while in region D are blueshifted, which is suggestive of a bipolar outflow.  
Other velocity maps also support this interpretation.
Considering the strong interaction between NGC 5195 and NGC 5194, the northern ionized gas structure can also be produced by tidal stripping.
The highest velocity gas is seen in region A with the median velocity of [O~{\sc{iii}}] emission line reaching $\sim 270\,\rm km\, s^{-1}$. 
The median velocities of [O~{\sc{iii}}] in the region B, C and D are $\sim 170\,\rm km\, s^{-1}$, $\sim 100\,\rm km\, s^{-1}$ and $\sim -50\,\rm km\, s^{-1}$, respectively. 
The [N~{\sc{ii}}] and H$\alpha$ velocity maps also exhibit similar kinematics as the [O~{\sc{iii}}] map in these four regions.

\subsection{Emission-line diagnostic diagram and map} \label{subsec:eline-map}

\begin{figure*}[ht!]
\includegraphics[width=\textwidth,trim=0 0 0 0]{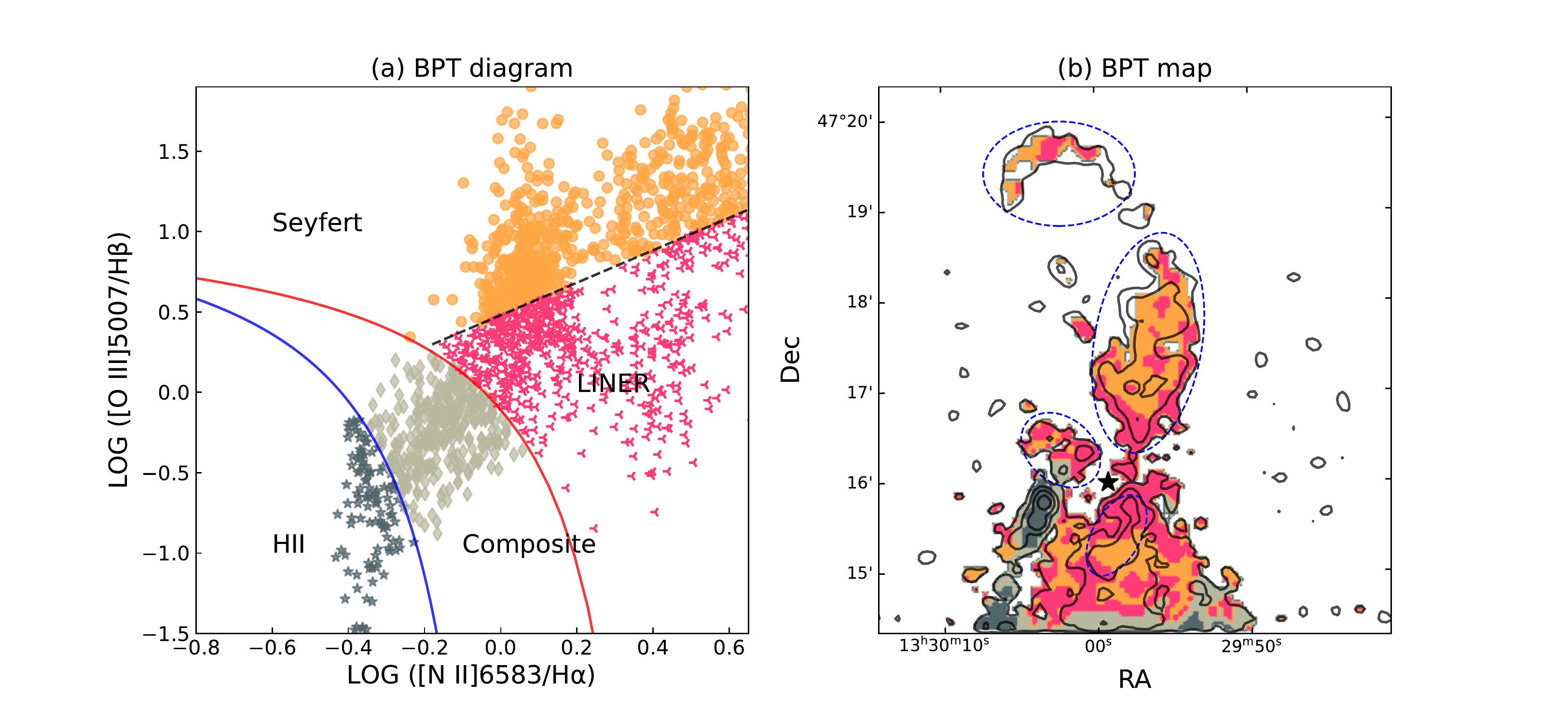}
\caption{(a): The BPT diagnostic diagram. The blue solid line is the classification boundary in \citet{2003MNRAS.346.1055K} for finding AGN. The red solid line shows the theoretical upper limit for star-forming galaxies \citep{2001ApJ...556..121K}, which separates the composite emission and AGN/LINER. The dashed line \citep{2010MNRAS.403.1036C} shows the division between AGN and LINER.
(b): Spatially resolved BPT map. The colors denote different ionization mechanisms in (a).
}
\label{fig:bpt}
\end{figure*}

The Baldwin, Phillips, and Telervich (BPT) diagnostic diagram \citep{1981PASP...93....5B} is usually used to distinguish the ionization and excitation mechanisms of the ionized gas.  
To obtain statistically robust results, the spaxels with the flux of the [O~{\sc{iii}}]$\lambda5007$ ($F_{\rm [O III]}$) or the [N~{\sc{ii}}]$\lambda6583$ ($F_{\rm [N II]})$ lower than $2\times10^{-17}\rm\, erg\,cm^{-2}\,s^{-1}$ are masked out in the diagram and map.
For spaxels with the flux of H$\alpha$ of $F_{\rm H\alpha}<10^{-17}\rm\, erg\,cm^{-2}\,s^{-1}$, upper limits of Balmer emission lines of $F_{\rm H\alpha}=10^{-17}\rm\, erg\,cm^{-2}\,s^{-1}$ and $F_{\rm H\beta}=F_{\rm H\alpha}/3.1$ are given.

In Figure~\ref{fig:bpt}, the BPT diagram and map are presented.
The ionization sources in NGC 5195 are dominated by AGN and LINER. 
Region A and B are dominated by AGN photoionization, only the edges of region B are classified as LINER.
In region D, the LINER dominates the area near the nucleus, while the AGN becomes the main source of ionization with the increasing distance to the nucleus.

The spiral arm under region C, which is from NGC 5194, is dominated by regions classified as HII and composite.
Considering the bottom of the field is dominated by composite regions and the lower part of the FoV is located at a spiral arm of NGC 5194 (Figure~\ref{fig:df2_spec}), most of the ionized gas here should be from the NGC 5194, which is consistent with our initial suggestion in Section~\ref{subsec:flux-vel-maps}.

\subsection{Comparison with the radio and X-ray data} \label{subsec:multiwav}

\begin{figure*}
\includegraphics[width=\textwidth,trim=0 70 0 70]{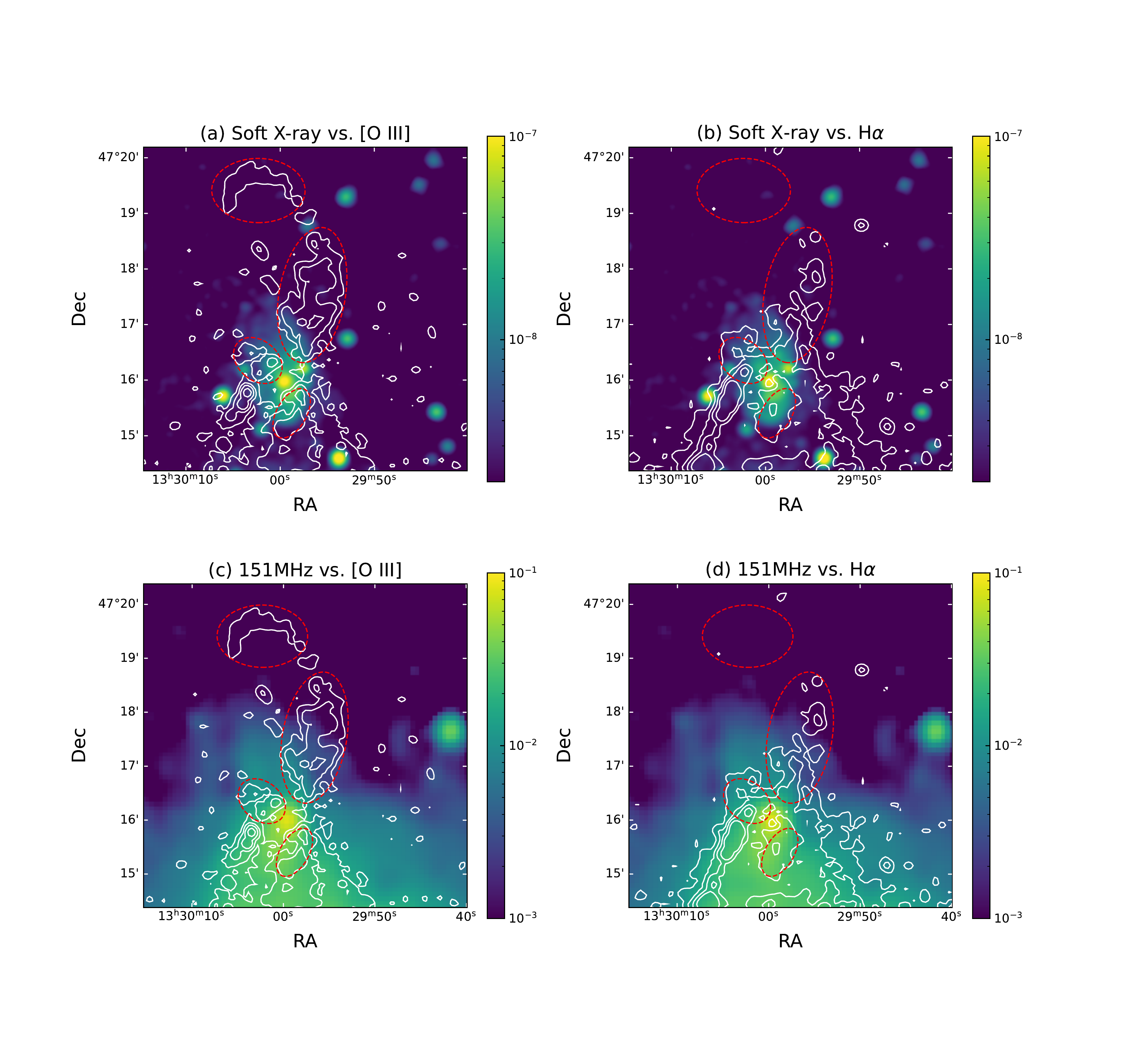}
\caption{(a): \textit{Chandra} soft X-ray (0.4--$2.0\rm\, keV$) image superposed with the contours of [O~{\sc{iii}}]$\lambda$5007 flux.
The unit of the colorbar is $\rm counts\,cm^{-2}\,s^{-1}$.
(b): Same as (a) but superposed with the contours of H$\alpha$ emission.
(c): LOFAR $151\rm\,MHz$ image \citep[][]{2014A&A...568A..74M} superposed with the contours of [O~{\sc{iii}}]$\lambda$5007 flux.
The unit of the colorbar is $\rm Jy/beam$.
(d): Same as (c) but superposed with the contours of H$\alpha$ emission.
The red dashed elliptical regions are the same as those in Figure~\ref{fig:fmap}.
}
\label{fig:sxay_lofar}
\end{figure*}

Figure~\ref{fig:sxay_lofar}(a) and (b) show the adaptively smoothed \textit{Chandra} soft X-ray (0.4--$2.0\rm\, keV$) image superposed with contours of [O~{\sc{iii}}]$\lambda$5007 and H$\alpha$ emission, respectively. 
The southern double-arc-like X-ray structure found by \cite{2016ApJ...823...75S} is spatially coincident with the inner part of [O~{\sc{iii}}] arc (region D), whereas the northern extended X-ray feature is inside the rims of the northern [O~{\sc{iii}}] structure (region B and C).
Compared to the H$\alpha$ flux contours, The soft X-ray emission also corresponds to the diffuse H$\alpha$ structure spatially, which is consistent with previous works \citep[][]{2016ApJ...823...75S,2018MNRAS.476.2876R}.

The Low Frequency Array (LOFAR) $151\rm\,MHz$ images \citep[][]{2014A&A...568A..74M} superposed with contours of [O~{\sc{iii}}]$\lambda$5007 and H$\alpha$ are presented in Figure~\ref{fig:sxay_lofar}(c) and (d), respectively.
As Figure~\ref{fig:sxay_lofar}(c) shows, a radio lobe or bubble extends from the nucleus to the northeast and just inside the northern [O~{\sc{iii}}] emission, indicating a likely association with a jet or outflow.
In Figure~\ref{fig:sxay_lofar}(d), the northern H$\alpha$ structure is similar to [O~{\sc{iii}}] but extends to a smaller scale.
The low-frequency radio extension is consistent with the northern radio bubble detected by \cite{2018MNRAS.476.2876R} using VLA, although that bubble-like feature is weaker with a smaller scale.
Note that \cite{2018MNRAS.476.2876R} also detected a radio bubble in L, C, and X bands in the southern region, which is spatially associated with the southern H$\alpha$, [O~{\sc{iii}}], and soft X-ray emission features.

\section{Discussion} \label{sec:discussion}

\subsection{Origin of the large-scale ionized gas structure} \label{subsec:origin}

As the flux and velocity maps show (Figure~\ref{fig:fmap} and \ref{fig:vmap}), the main ionized gas structure in NGC 5195 (regions A, B, C, and D) closely resembles a bipolar outflow.
According to the bubble-like morphology detected in different bands ([O~{\sc{iii}}], H$\alpha$, [N~{\sc{ii}}], and the soft X-ray) and the corresponding LOFAR radio bubble feature (Figure~\ref{fig:sxay_lofar}), we suggest that this large-scale ionized structure can be associated with an AGN jet or outflow, which is consistent with \cite{2018MNRAS.476.2876R} but on a larger scale.
The BPT diagram and map (Figure~\ref{fig:bpt}) also suggest that the main ionized structures (regions A, B, and D) should be photoionized by an AGN.
The LINER regions in the BPT map spatially correspond to the superposed areas of the soft X-ray, radio, and ionized gas emission, which indicates these regions can be excited by the shocks produced by a jet or outflow.
Considering the current low state of the SMBH in the center of NGC 5195 \citep[][]{2018MNRAS.476.2876R} and the weak [O~{\sc{iii}}] flux in the nuclear region, the large-scale ionized gas structure should relate to past epochs of AGN activities, which is consistent with the suggestion in previous works \citep[][]{2016ApJ...823...75S,2018MNRAS.476.2876R}.

The dynamical timescale of the ionized outflow can be estimated to be $t_{\rm dyn} \approx D_{\rm out}/v_{\rm out}=30\rm\, Myr$, where $D_{\rm out}\sim10\rm\,kpc$ denotes the distance away from the nucleus and $v_{\rm out}\sim300\rm\,km\,s^{-1}$ is the outflow velocity.
The simulation expects that the recent interaction between NGC 5195 and NGC 5194 has occurred $\approx50$--$100\rm\,Myr$ ago \citep[][]{2000MNRAS.319..377S}.
Considering the gas infalling timescale \citep[][]{2016ApJ...823...75S}, the dynamical timescale of the ionized outflow is roughly corresponding to the recent interaction between NGC 5195 and NGC 5194.

Distant ionized clouds originated from tidal stripping and illuminated by past AGN activities have been detected in some galaxies \citep[e.g.][]{2017ApJ...835..256K}, and such a connection appears to be quite common as revealed by systematic surveys \citep{2022MNRAS.510.4608K}.
The Hanny’s Voorwerp is a famous example extending $\sim 50\,\rm kpc$ in projection from the nucleus of IC 2497, which is inferred as a part of massive HI tidal tail and photoionized by past luminous AGN \citep[][]{2009A&A...500L..33J,2009MNRAS.399..129L,2012AJ....144...66K}.
In the M51 system, large-scale neutral gas structures originating from tidal stripping have been revealed by \cite{1990AJ....100..387R} using VLA.
The large ionized gas structure (regions A and B) is coincident with a part of a giant tidal tail traced by HI emission extending from NGC 5195 to the north.
Considering the velocity of the corresponding HI tidal feature is about 30--$90\rm\,km\,s^{-1}$ \citep[][]{1990AJ....100..387R}, which is several times lower than the median velocity ($\sim 170$--$270\,\rm km\, s^{-1}$) of the northern [O~{\sc{iii}}] structure, the ionized gas feature may be a shocked, relatively higher velocity component of the tidal stream or photoionized and accelerated by a more active AGN in the past.
Combining the dominant AGN photoionization mechanism in this ionized structure suggested by the BPT diagram, the spatial correspondence with the HI tidal tail and the faded AGN in the nucleus of NGC 5195, the northern large-scale ionized gas structure is similar to Hanny’s Voorwerp, which can be a part of tidal tail photoionized by past AGN activities.

In fact, tidal stripping and AGN activities can both be important here. 
In such complicated interacting galaxy pairs, it is challenging to pinpoint the origin of the large-scale ionized gas structure. 
More sensitive observations on the faint structures are needed to resolve this issue.

\subsection{Energetics of the ionized outflow} \label{subsec:energetics}

Assuming the ionized gas structure in NGC 5195 is a biconical outflow, we can estimate the mass and energy outflow rate.
The mass of an ionized outflow can be estimated using H$\beta$ emission line \citep[e.g.][]{2014MNRAS.441.3306H} or [O~{\sc{iii}}]$\lambda5007$ line \citep[e.g.][]{2015A&A...580A.102C,2016ApJ...833..171K}.
We firstly use the [O~{\sc{iii}}]$\lambda5007$ line to calculate the outflow mass following \citep[e.g.][]{2015A&A...580A.102C}:
\begin{equation}
	M_{\rm gas}=4\times10^{8} (\dfrac{L_{\rm [O III]}}{10^{44}\rm\ erg\ s^{-1}})(\dfrac{100\rm\ cm^{-3}}{n_{e}})\rm\ M_{\astrosun}, 
	\label{eq:1}
\end{equation}
where $L_{\rm [O III]}$ is the luminosity of [O~{\sc{iii}}]$\lambda5007$ and $n_{\rm e}$ is the electron density, assuming the solar metallicity.
The electron density can be estimated from the flux ratio of [S~{\sc{ii}}]$\lambda 6716/$[S~{\sc{ii}}]$\lambda 6731$, assuming the case B and the electron temperature $T_{e} = 10^{4}$ K \citep[][]{2006agna.book.....O}.
However, the [S~{\sc{ii}}] doublets are very weak in most areas of our datacube, especially in regions A, B, and C.
In region D, the [S~{\sc{ii}}] doublets can be fitted but with $\rm SNR<3$, and the corresponding $n_{\rm e}$ is estimated to be a very low value ($<1 \rm\, cm^{-3}$), which is much lower than the sensitive range ($10<n_{\rm e}<10^{4}\rm\, cm^{-3}$) of these doublets \citep[][]{2006agna.book.....O}.
\cite{2018MNRAS.476.2876R} estimated the density to be $n_{\rm e}\approx 0.01 \rm\, cm^{-3}$ following the method described by \cite{2006IAUS..230..293P}, which is consistent with the very low value indicated by the flux ratio of the [S~{\sc{ii}}] doublets.
The mass and energy outflow rate can be calculated following:
\begin{align}
    \dot{M}_{\rm out} &= 3M_{\rm gas}\dfrac{v_{\rm out}}{R_{\rm out}}, \label{eq:2}\\
    \dot{E}_{\rm out} &= \dfrac{1}{2}\dot{M}_{\rm out}v_{\rm out}^{2}, \label{eq:3}
\end{align}
where $v_{\rm out}$ denotes the averaged velocity of the outflow, and $R_{\rm out}$ is the outflow size.
Only the luminosity of [O~{\sc{iii}}] from regions A, B, C, and D is considered in estimating the outflow rates.
Adopting $n_{\rm e}= 0.01 \rm\, cm^{-3}$, $R_{\rm out} = 10 \rm\, kpc$, and $v_{\rm out} = 300 \rm\, km\,s^{-1}$, we obtain $\dot{M}_{\rm out} = 3.5 \rm\, M_{\astrosun}\, yr^{-1}$ and $\dot{E}_{\rm out} = 9.8\times10^{40}\rm\, erg\, s^{-1}$.
This estimated $\dot{E}_{\rm out}$ is lower than but consistent with the value (a few $10^{41}\rm\,erg\,s^{-1}$) calculated by \cite{2018MNRAS.476.2876R}, because the outflow mass estimated from the [O~{\sc{iii}}]$\lambda5007$ flux is normally a lower limit \citep[][]{2015A&A...580A.102C}.

The upper limit of outflow mass can be estimated using the total H$\beta$ luminosity of NGC 5195, which is derived using $L_{\rm H\beta} \sim L_{\rm H\alpha}/3.1 = 4.5\times10^{38}\rm\,erg\, s^{-1}$.
The outflow mass traced by H$\beta$ emission line can be calculated following \citep[e.g.][]{2014MNRAS.441.3306H,2022ApJ...933..110X}:
\begin{equation}
	M_{\rm gas}=6.78\times10^{8} (\dfrac{L_{\rm H\beta}}{10^{43}\rm\ erg\ s^{-1}})(\dfrac{100\rm\ cm^{-3}}{n_{e}})\ M_{\astrosun}. \label{eq:4}
\end{equation}
We obtain the upper limits of the mass and energy outflow rate of $\dot{M}_{\rm out,max} = 27.9 \rm\, M_{\astrosun}\, yr^{-1}$ and $\dot{E}_{\rm out,max} = 7.9\times10^{41}\rm\, erg\, s^{-1}$, respectively, which are nearly one order of magnitude larger than the lower limits.

Comparing with the kinetic power resulting from the stellar winds and SNe \citep[][]{2018MNRAS.476.2876R} given the current $\rm SFR\approx 0.1\,M_{\astrosun}\, yr^{-1}$ of NGC 5195 \citep[][]{2016ApJ...830..137A}, even the lower limit of the energy outflow rate is larger by a factor of two to three, which imply that the current star formation process cannot drive the ionized outflow.
A starburst that occurred 370--480 Myr ago had been suggested by \cite{2012ApJ...755..165M}, while this is not consistent with the timescale of the ionized outflow ($t_{\rm dyn} \approx 30\rm\, Myr$).

\begin{figure*}
\includegraphics[width=1.1\textwidth,trim=0 0 0 0]{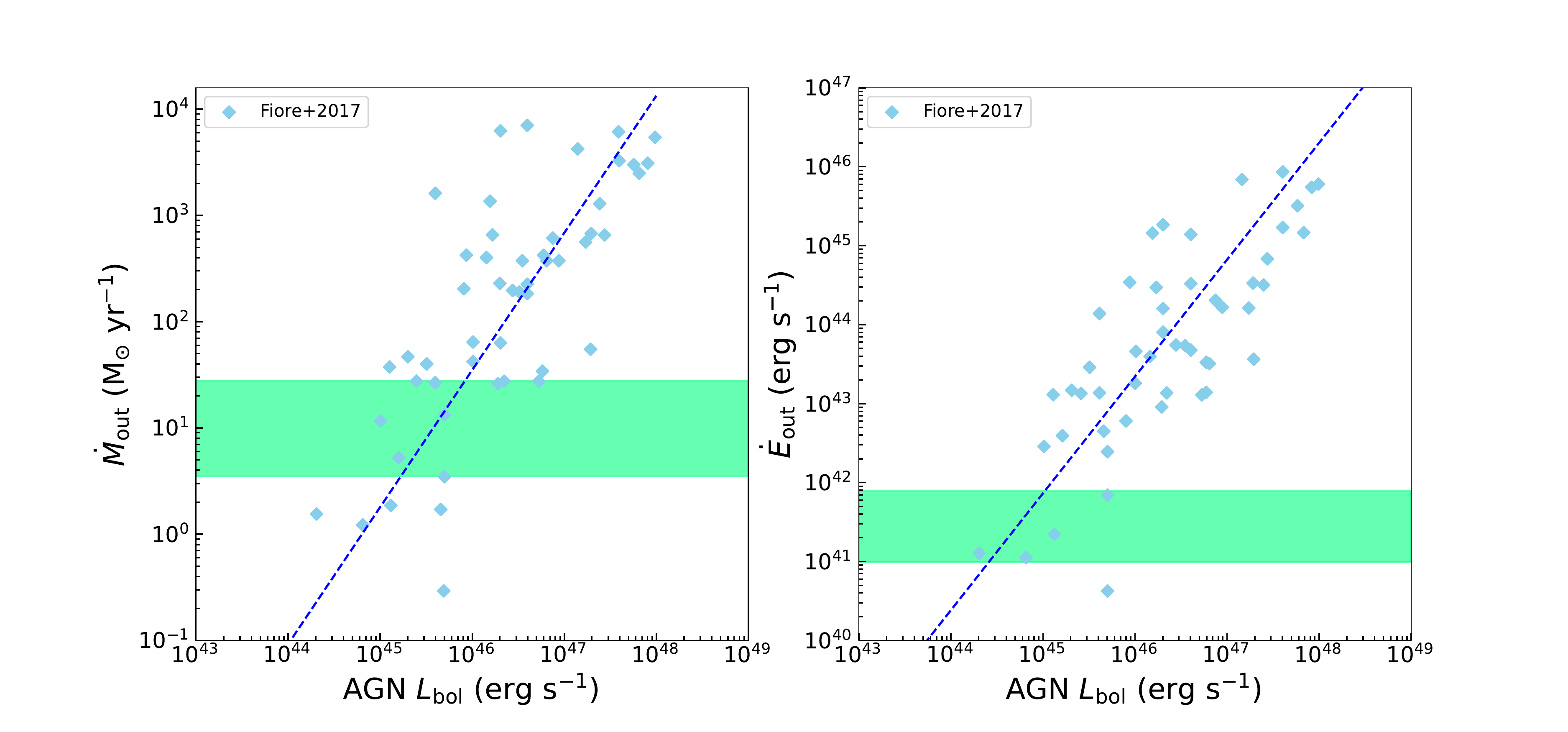}
\caption{Mass (left panel) and energy (right panel) outflow rate as a function of the AGN bolometric luminosity. 
The lightblue diamonds denote the ionized outflow measurements from \citet{2017A&A...601A.143F} and the blue dashed line is their best-fit correlation. 
The green regions show lower and upper limits of the outflow rates of NGC 5195.}
\label{fig:F17correl}
\end{figure*}

To compare the outflow rates of NGC 5195 with observational results of other AGNs, we show the outflow rates as a function of the AGN bolometric luminosity following \citet{2017A&A...601A.143F} in the Figure~\ref{fig:F17correl}.
The green regions in the two panels denote the ranges of the outflow rates derived above.
The outflow rates of NGC 5195 are comparable to the lower part of the correlations. 
If the ionized outflow of NGC 5195 is radiatively driven by the AGN and follows the correlation of \citet{2017A&A...601A.143F}, an AGN bolometric luminosity of $L_{\rm bol}\sim 10^{44.4}$--$10^{45.9}\rm\, erg\,s^{-1}$ is inferred, which is approaching or exceed the Eddington luminosity of $L_{\rm Edd}\sim 1.3\times 10^{45}\rm\, erg\,s^{-1}$, adopting $M_{\rm BH}\sim 10^{7}\rm\,M_{\astrosun}$ for the mass of the SMBH.  
Alternatively, if the ionized outflow is driven by an AGN jet \citep[][]{2018MNRAS.476.2876R}, the outflow rate can be much higher than expected from the correlation of \citet{2017A&A...601A.143F} for a low luminosity AGN. In either case, the outflow is likely powered by the past powerful AGN activities, given that the AGN in NGC 5195 is currently very weak with a small-scale radio jet \citep[e.g.][]{2016ApJ...823...75S,2018MNRAS.476.2876R}.

Combining the large size and the relatively high outflow rates of this bipolar outflow, feedback effects might be significant in this galaxy.

\section{Conclusions} \label{sec:summary}

In this work, we report the observation of the nearby interacting galaxy NGC 5195 using the CFHT/SITELLE, and study the spatially resolved ionized gas.  
To identify the ionization mechanism of the emission line gas structure, the BPT diagnostic diagram and map are derived.
Multiwavelength observations including the soft X-ray (0.4--$2.0\rm\,keV$) data from the \textit{Chandra} archive and the LOFAR $151\rm\,MHz$ image from \cite{2014A&A...568A..74M} are used to compare with our spatially resolved data.
The main results and conclusions are summarized as follows:

\begin{enumerate}
    \item A remarkable $\sim 10$ kpc scale (in projection) ionized structure traced by [O~{\sc{iii}}], [N~{\sc{ii}}], and H$\alpha$ (less extended) emission lines is discovered (Figure~\ref{fig:fmap} and Figure~\ref{fig:vmap}).
    
    \item According to the BPT diagram and map, the large-scale structure is dominated by the region classified as Seyfert and LINER (Figure~\ref{fig:bpt}), which indicates that the structure is likely associated with AGN activities or shocks.
    
    \item The northern soft X-ray and the LOFAR $151\rm\,MHz$ bubble-like feature is just located inside the ionized gas structure, which strongly indicates that the ionized gas feature is associated with an AGN jet or outflow (Figure~\ref{fig:sxay_lofar}).
    Considering the strong interaction in the M51 system, the northern large-scale ionized gas structure may have a complicated origin. 
    This structure can be a part of the giant HI tidal tail revealed by previous works \citep[][]{1990AJ....100..387R} and photoionized by past AGN activities. 
    
    \item Assuming the ionized gas structure is a bipolar outflow, the mass and energy outflow rates can be estimated to be $\dot{M}_{\rm out} = 3.5$--$27.9 \rm\, M_{\astrosun}\, yr^{-1}$ and $\dot{E}_{\rm out} = 0.98$--$7.9\times10^{41}\rm\, erg\, s^{-1}$, which cannot be provided by the current star formation activities (Section~\ref{subsec:energetics}).
    In lieu of the currently weak AGN, the ionized outflow is likely associated with the past powerful AGN activities.
    
\end{enumerate}

The NGC 5195 galaxy could be the closest example exhibiting a projected $\sim 10$ kpc scale ionized structure associated with past AGN outflow/jet activities or light echoes, which makes it an ideal laboratory for studying the AGN feedback, past AGN activities, and the interaction between the galaxy pair. 
Our future work will study the M51 system in more detail deploying the rich archival multi-wavelength data.

\section{Acknowledgments}

We thank the anonymous referee for helpful comments that significantly improved the clarity of our work. 
We thank Thomas Martin for the instructions on the SITELLE data analysis. J.W. acknowledges the NSFC grants U1831205, 12033004, 12221003 and the science research grants from CMS-CSST-2021-A06 and CMS-CSST-2021-B02. Our SITELLE observation (19AS002) is kindly supported by China Telescope Access Program (TAP). Based on observations obtained at the Canada-France-Hawaii Telescope (CFHT) which is operated from the summit of Maunakea by the National Research Council of Canada, the Institut National des Sciences de l'Univers of the Centre National de la Recherche Scientifique of France, and the University of Hawaii. The observations at the Canada-France-Hawaii Telescope were performed with care and respect from the summit of Maunakea which is a significant cultural and historic site. Based on observations obtained with SITELLE, a joint project between Universit$\acute{\rm e}$ Laval, ABB-Bomem, Universit$\acute{\rm e}$ de Montr$\acute{\rm e}$al and the CFHT with funding support from the Canada Foundation for Innovation (CFI), the National Sciences and Engineering Research Council of Canada (NSERC), Fond de Recheche du Qu$\acute{\rm e}$bec - Nature et Technologies (FRQNT) and CFHT.
This research has made use of data obtained from the Chandra Data Archive and the Chandra Source Catalog, and software provided by the Chandra X-ray Center (CXC) in the application packages CIAO and Sherpa.


%

\vspace{5mm}
\facilities{CFHT (SITELLE), CXO (ACIS), LOFAR}


\software{astropy \citep{2013A&A...558A..33A,2018AJ....156..123A}, CIAO \citep{2006SPIE.6270E..1VF}, DS9 \citep{2003ASPC..295..489J}, ORCS \citep{2015ASPC..495..327M}
          }

\bibliography{5195}{}

\bibliographystyle{aasjournal}


\end{CJK*}
\end{document}